# Self-repairing hardware architecture for safety-critical cyber-physical-systems

*Shawkat S. Khairullah*[1] ✉, *Carl R. Elks*[2]
[1]Department of Computer Engineering, University of Mosul, Mosul, Iraq
[2]Department of Electrical and Computer Engineering, Virginia Commonwealth University, 601 West Main Street, Richmond, USA
✉ E-mail: khairullahss@mymail.vcu.edu

**Abstract:** Digital embedded systems in safety-critical cyber-physical-systems (CPSs) require high levels of resilience and robustness against different fault classes. In recent years, self-healing concepts based on biological physiology have received attention for the design and implementation of reliable systems. However, many of these approaches have not been architected from the outset with safety in mind, nor have they been targeted for the safety-related automation industry where the significant need exists. This study presents a new self-healing hardware architecture inspired by integrating biological concepts, fault tolerance techniques, and IEC 61131-3 operational schematics to facilitate adaption in automation and critical infrastructure. The proposed architecture is organised in two levels: the critical functions layer used for providing the intended service of the application and the healing layer that continuously monitors the correct execution of that application and generates health syndromes to heal any failure occurrence inside the functions layer. Finally, two industrial applications have been mapped on this architecture to date, and the authors believe the nexus of its concepts can positively impact the next generation of critical CPSs in industrial automation.

## 1 Introduction

As we move toward a universe of networkable objects, cyber-physical production systems, distributed manufacturing, and distributed automation emerges as a set of loosely coupled, smart, autonomous units working together to achieve plant operations in ways we could not imagine a few decades ago. To fully achieve the promise of autonomic behaviour and resiliency in advanced automation, these cyber-physical systems (CPSs) will need to rely on controllers that have properties of resilience, self-healing, and agility. System resilience is 'the ability of organisational, hardware and software systems to mitigate the severity and likelihood of failures or losses, to adopt to changing conditions, and to respond appropriately after the fact' [1]. System resilience and autonomic computing methods are emergent technologies organised around a concept of self-governance – much in the way biological systems have evolved. From the broadest stance, system resilience can be seen as four interwoven characteristics: self-configuration, self-optimisation, self-healing, and self-protection. Of these four attributes, self-healing and self-protection capabilities are necessary requirements with respect to *critical* CPSs; where production downtime, damage to expensive machinery, and serious injury can occur [2, 3]. Over the last two decades, new research on self-healing digital systems inspired by biology has been growing steadily – to enhance fault tolerance, resilience, and survivability properties [4–10]. To date, most of these approaches have not addressed automation or Instrumentation and Control (I&C) applications. Specifically, we found the 'programmability' aspects of the previous designs to be too low level for application engineers to understand (e.g. VHDL and bit-level implementations), or they lacked the computational capacity required for control applications (i.e. simple logic). To address these issues, among others, this paper presents a new self-healing hardware architecture with its main fourfold contributions: (i) Achieving high levels of tolerance against different classes of failures and faults. (ii) Providing multiple layers of defense, healing, and graceful reconfiguration. (iii) Manage complexity to enhance verification and validation (V&V) properties. (iv) And providing programmability via function blocks constructs – to enhance accessibility for practicing engineers.

Our inspiration for this research is not aimed toward blindly mimicking these processes, but motivated by comprehending their robust operation and using it as a foundation to build a new self-healing architecture with its concept that leverages existing programmable and configurable hardware technologies such as field-programmable gate array (FPGA) technology, and application-specific integrated circuits (ASICs). Many studies (and marketplace products) have confirmed that FPGA technology is a viable and competitive option for various application domains: aerospace, nuclear industry, and industrial control systems due to their characteristics with regard to re-programmability, high performance, concurrency, and low-cost development. The other main advantage of using FPGA technology is the large ecosystem of design, programming, and verification tools that allow users to build custom designs directly onto FPGAs [11, 12].

Aggressive technology and power scaling have, over the past few decades, led to nano-electronic devices that are faster, more power-efficient and cheaper with each new generation. However, the reliability of highly scaled digital systems has been on a gradual decline for the past decade. This can be attributed in part to the increased sensitivity of highly scaled devices to transient faults arising from random events such as high-energy particle strikes, high-intensity fields, electrostatic discharge, and power fluctuations. Accordingly, contemporary IC technology (45 nm and below) devices tend to manifest more single-event upset (SEU) and multi-event upset (MEU) and transient fault occurrences than previous technology generations [13, 14].

The proposed architecture in this paper is guided by a combination of two design principles: biologically inspired concepts and architectural principles. First, we present the biological techniques: the cell life cycle and the immune system used by living organisms to achieve self-diagnostic and self-healing against the invaders. The cell life cycle is considered to be a self-diagnostic mechanism used to make the cell capable of continuously monitoring its chemical internal state through four phases of testing integrity. Furthermore, if the living cell is attacked by an invader causing a weak mutation, the effects of this mutation will be tolerated using a special enzyme. However, if the mutation is strong that cannot be repaired inside the cell and it leads to a cell death, the immune system will start working by generating two types of stem cells: adult cells, B-cells and T-cells which, are divided to produce more stem cells (self-renewal), and embryonic cells that can be divided in order to generate terminally differentiated cells [15].

Second, architectural principles establish the basis of the formulation of operational rules for organisation, and rules for composition for any architecture. Jackson [16] lists four categories of biologically inspired resilience attributes, namely capacity, flexibility, tolerance, and inter-element collaboration to achieve resilience in architecting systems We extend the aforementioned attributes with three additional principles:

(i) *Reorganisation*. Flexibility is the ability of a system to undergo changes with relative ease in operation while experiencing faults and disturbances. In contrast, the reorganisation principle says that the system should be able to restructure itself and/or migrate functionality in response to disruptions.

(ii) *Separation of concerns or partitioning*. Building self-contained units allow the disentangling of separate functions. In turn, they can be grouped in self-contained architectural units to generate stable forms. Partitioning in the proposed architecture separates the healing layers and computational activities from communication and I/O activities.

(iii) *Independence*. Reducing the architectural units to their minimum representation as required by the application.

The design principles listed above are used throughout the design of the proposed bio-inspired digital system to achieve high levels of self-healing capacity and architect an efficient self-healing, engineer accessible hardware architecture.

## 2 Related work

Over the last two or three decades, new research on the design and realisation of self-healing digital systems inspired by biological concepts has been steadily growing to enhance dependability properties of safety-critical applications against different classes of faults: transient, intermittent, and permanent. In this section, a brief discussion on bio-inspired hardware-based digital systems that can be realised on FPGA fabric is presented. A relatively new emerging field that is closely related to our research project for the realisation of resilient digital systems is bio-inspired system design. It attempts to go beyond traditional approaches of fault-tolerant computing and modular redundancy to learn from characteristics of living things and adapting them to digital electronic systems [17, 18]. This research topical area aims to achieve high levels of resilience and self-healing properties by utilising the power of reconfigurable hardware computing. The self-healing mechanism can be partitioned basically into five stages: fault model stage, fault detection stage, faulty component isolation stage, system reconfiguration stage, and system self-healing stage. In [19, 20], a new programmable cellular architecture that performs logic and arithmetic operations for a self-repairing FPGA has been designed by Mange *et al.* This architecture includes four hierarchal levels of organisation: molecular, cellular, organismic, and population to tolerate the transient faults in the molecular layer. In [4], Ortega and Tyrrell have presented a different architecture which embeds a logic block performing the functions by a 2-1 multiplexer and a D flip-flop. This approach is a two-level hierarchical architecture consisting of a cellular level and organism level. Two modes of operation were used in this work to control the operation of the organism in a fault-tolerant manner. Zhang *et al.* [5] have developed an architecture that also works at two levels: cellular and organism, but they have increased the level of fault coverage by adding more levels of reliability to detect the transient faults in the configuration memory besides the detection of permanent faults. In [6], the authors have designed a new architecture inspired by the biological process of the human immune system. Their approach used several routing cells and spare cells distributed among an array of functional cells. In [21], Wang and co-authors have designed a different approach to realise the self-healing cellular architecture. Their design was based on using a lookup table (LUT) as a building block for the functional unit. In [7], a novel self-repairing hardware architecture inspired by paralogous gene regulatory circuits was designed to achieve fast fault recovery with efficient use of hardware resources. Ultimately, in [22, 23] the researchers have developed a new self-healing HW architecture which has some similarities with PLC based architecture (e.g. use of function blocks for programming, hardware/FPGA based), the proposed architecture is a completely bio-inspired architectural approach which includes fault aware function blocks, reconfiguration ability, and the ability to migrate functionality.

Most of the previous related designs presented in this section have focused on the latter two stages of the self-healing mechanism, in particular, stage 4 and stage 5, which include the reconfiguration strategy of the system and how its structure is reconfigured so that the system can adapt to the fault occurrence. However, our aim in this research is to design a new self-healing hardware architecture that is inherently resilient, capable of being verified for safety properties, and accessible by the industrial automation. This architecture is aimed to provide resilience properties to multiple classes of faults and verify safety and functional properties for the critical aspects of the design using concepts of formal design assurance to an effort to qualify the novel self-healing system to be used in safety-critical applications.

## 3 Expected contributions to the field

The research on bio-inspired self-healing digital systems is expected to be noteworthy to a number of stakeholders in the extreme environment digital Instrumentation and Control (I&C) systems, industrial automation, and those areas concerned with safety-related CPSs. As these different applications domain areas become pervasive, more CPS applications will be deemed critical for public services – such as smart traffic automation, smart energy management, and smart cities. Digital embedded devices operating in these diverse application areas may experience harsh operating conditions and environmental changes where disturbances from random events such as high-intensity radiated electromagnetic (EM) fields (HIRFs), extreme temperatures, radiation, or cosmic particle strikes are at an increased threat. In all these cases, the occurrence of transient and permanent faults occurring, simultaneously affecting digital embedded devices or nodes, is a significant concern. As a consequence, the ability to detect and repair the experienced failure modes is important. In order to support all the objectives mentioned above, the work presented in this paper can be categorised in three contributions as described below:

(i) This research extends the state-of-the-art resilient very-large-scale integration (VLSI) design by developing new Fault Management Approach to support resiliency not previously reported in the literature. To the best of our knowledge, the majority of work on bio-inspired digital systems achieve the self-healing objective at decentralised level by embedding self-diagnostic modules inside the functional cells. These models are used to configure an internal control unit or notify a neighbouring spare cell as a recovery mechanism. In our approach, we instead have health monitoring and recovery units.

(ii) To date, most of the traditional biologically-inspired self-healing approaches have not seriously addressed industrial automation or safety-related I&C applications. Specifically, we found the 'programmability' aspects of the previous designs to be too low level for application engineers to understand (e.g. very-high-speed integrated circuit hardware description language (VHDL) and bit-level implementations), or they lacked the computing capacity required for control applications (i.e. simple logic). For example, as a result, a unique hierarchal self-healing architecture is designed in that resilience principles are derived from a heterogeneous perspective-combining concepts from biological systems (immune system, stem cells, living cell cycle, and genetic expression) and computer organisation to provide a well-formed self-healing hardware architecture.

(iii) To the best of our knowledge, we believe this architecture is the first to employ PLC programming semantics accompanied by traditional fault tolerance techniques in a bio-inspired self-healing architecture. Programming semantics for PLC controllers typically specifies the standards for PLC software, and these standards can define the PLC configuration, programming, and data storage. PLC vendors typically specify five basic programming languages that support the IEC 61131 standard and these programming languages are Functional Block Diagram (FBD), Ladder Diagram (LD), Sequential Functional Chart (SFC), Structured Text (ST), and Instruction List (IL).

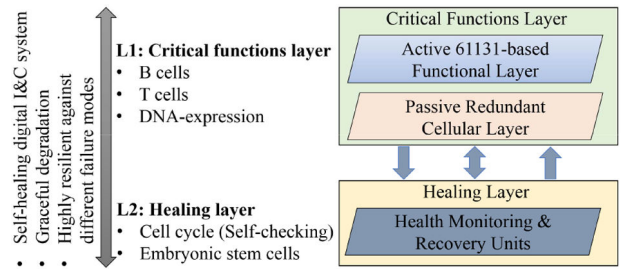

**Fig. 1** *High-level perspective of the proposed architecture*

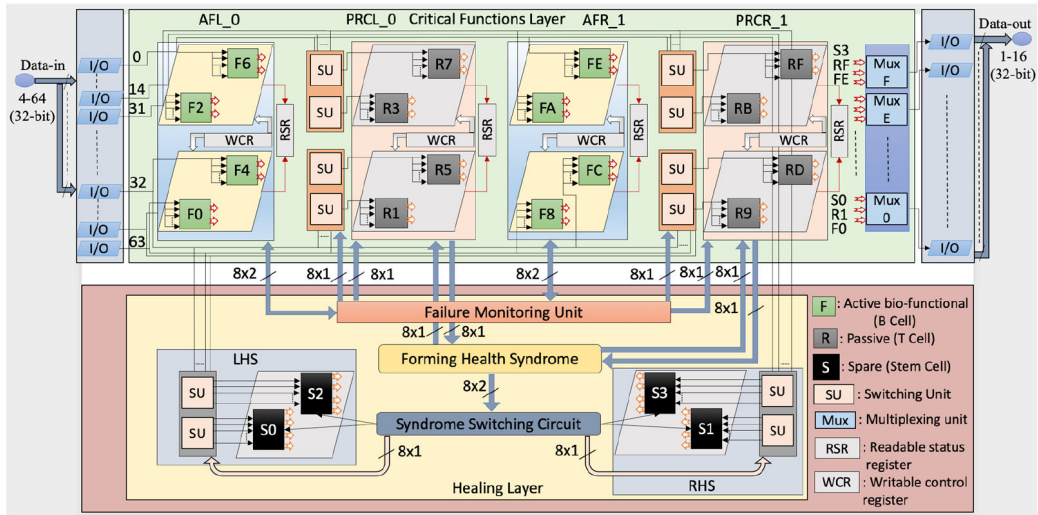

**Fig. 2** *Internal structure of the architectural concept*

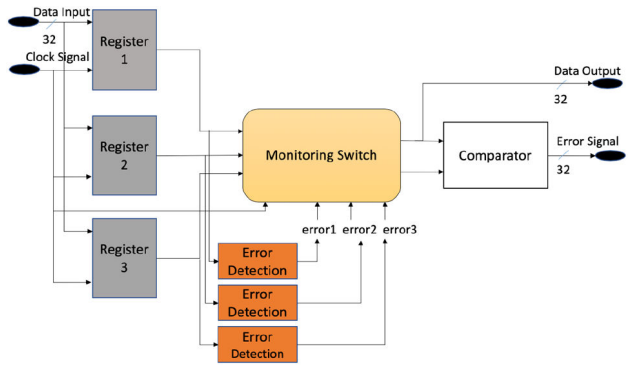

**Fig. 3** *Hybrid redundancy unit*

## 4 Overview of the proposed hardware architecture

The proposed architecture is designed to achieve high levels of self-healing capacity against different classes of faults (transient, permanent, and hardware common cause failures (CCFs)) with utilising a minimum amount of hardware resources. It also aims to provide efficient hardware area overhead, and verification and validation (V&V). The basic architecture shown in Fig. 1 is based in part on the way biological organisms achieve resiliency, and our architectural design principles. It is comprised of two principle divisions; (i) the Critical Functions Layer which is responsible for providing the intended functionality of critical application and (ii) the Healing Layer which is not only responsible for monitoring the healthy behaviour of the functions but also responsible for triggering the required recovery mechanisms to heal any defected T cells present in the critical layer. Furthermore, the critical layer embeds two sublayers: the active 61131-based functional sublayer (AFL) corresponding to B cells and the passive redundant cellular sublayer (PRCL) imitated the T cells in the immune system. However, the healing layer is corresponding to the life cycle of the living cell and the embryonic stem cells. We utilise these two main layers to create interacting functional and self-healing partitions to achieve overall system resilience (Fig. 2).

Additionally, the *critical functions layer* executes the safety-critical application functions. Specifically, it contains 16 functional cells: *eight* active B cells (designated *F* in Fig. 2) distributed among two active functions sublayers: left AFL and right AFR used to execute the application-based functions. The same layer also contains *eight* passive pre-generated redundant T cells connected as passive redundant resources: PRCL and PRCR used as a healing mechanism for the faulty B cells (designated *R* in Fig. 2).

The correct execution of each B cell is monitored continuously by its neighbouring healing layer, and once a fault is detected and determined to be transient inside the cell, it is masked/tolerated using an embedded hybrid redundancy unit. The hybrid redundancy unit (see Fig. 3) represents the first line of defense against the discovered transient faults defined as temporary deviations in the input register values. It is designed as an active redundancy technique to tolerate the effects of transient faults that may defect the input registers for the fault-tolerant generic function blocks (FTGFBs).

Each hybrid redundancy unit (see Fig. 3) consists of eight hardware components: three registers, three error detection units, a monitoring switch, and a comparator circuit. A transient fault was simultaneously injected into four input registers of the four hybrid redundancy units (HRUs) embedded inside the FTGFB in which transient faults can be tolerated sequentially for an unlimited number of times using a self-monitoring switching unit.

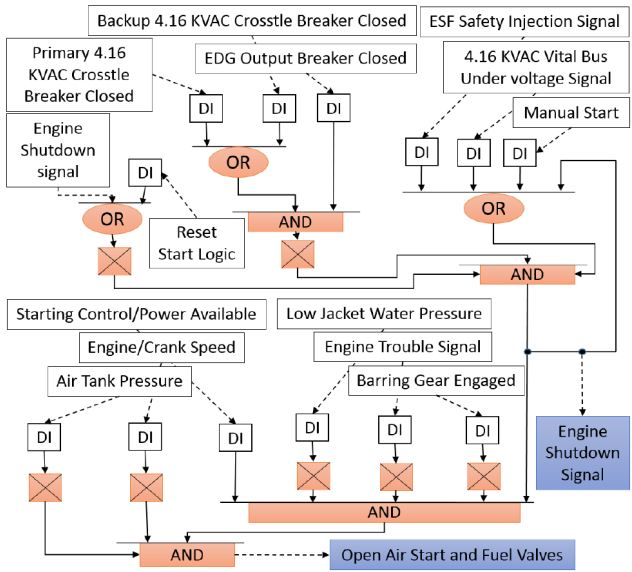

**Fig. 4** *Logic diagram for starting the EDG system*

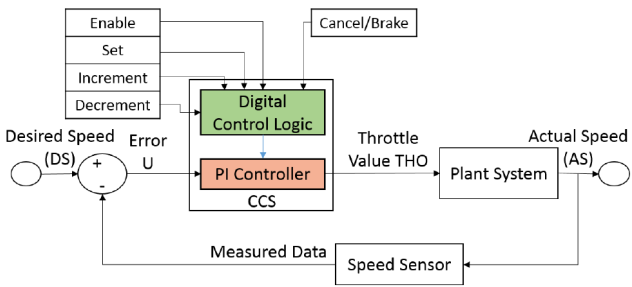

**Fig. 5** *Closed-loop cruise control system CCS*

Since this self-healing architecture is designed to realise the functionality of critical systems operating in harsh environments, radiation-induced transient faults that can occur at unpredictable times are the most prevalent fault type [24], and the process of hardening the digital device at the circuit level [25] is more effective, the subsequent fault-tolerance can occur even with an uncovered error. If they are not tolerated at the block level, their wrong values will be sensitised at the output signals of the cell level, which is directly connected to the external world through a network of I/O digital ports and can impact the safety of the public or the environment.

On the other hand, any permanent failure occurrence inside the functional B cells can be detected using a *passive duplication with a comparison unit* embedded inside the same cell, and an error flag is transmitted to the healing layer. Consequently, the *failure monitoring unit* embedded in the *healing layer* will sensitise that error and generates a health syndrome works as a self-healing mechanism. This mechanism includes sending one digital control signal to the defected AFL sublayer to deactivate the output of the faulty B cell (cell death). Another signal is transmitted to a routing unit that includes two switching units and is used to reroute digital input data from the faulty B cell and makes it available at the input of the healthy T cell (reorganisation). Finally, the third control signal selects a genetic code stored in a configuration memory of the T cell located in the neighbouring PRCL unit so that the functionality of the defected cell is healed and performed by this healthy T cell (restoration). As a result, the *failure monitoring unit* embedded in the healing layer represents the second line of defence against the permanent faults effecting one or all of the eight B cells.

As the third line of defence against the occurrence of additional permanent faults that can defect the T cells, four hardware components were added to the design of the healing layer, which is responsible for fault management for the entire critical functions layer. These components are (i) A forming health syndrome unit – which continuously reads the error signals from the two PRCL sublayers embedded inside the critical functions layer. Also, it generates eight syndromes used to differentiate eight 61131-based execution units embedded in four embryonic stem cells. (ii) A syndrome switching circuit – selects which one of the eight syndromes, generated by the health unit, will be chosen to differentiate the embryonic stem cell and its two embedded execution units. (iii) Two healing sublayers: LHS and RHS and each sublayer contain two embryonic stem cells (S0, S2 or S1, S3) – can be differentiated to repair any type of T cells in the critical layer. The healing layer fault management process is aimed at managing the processing capacity of the critical functions layer. If too many faults occur, then we have too few resources to maintain operations. In addition, this process is imitating how the embryonic stem cell is differentiated in the case of the failure of the immune system in generating the T cells as the first line of defence against the invader.

## 5 Critical applications applied to the proposed architecture

Two safety-critical cyber-physical applications have been mapped into the proposed architecture and these applications are Emergency Diesel Generator (EDG) Startup for Nuclear Power backup energy and an Automotive Cruise Control. These two applications are modest steps toward a planned at-scale-application related to complex distributed industrial control. Quartus Prime 15.1 Lite Edition software from Altera was used as a design tool that supports several FPGA device families and the system was embedded in a digital platform, the Altera Cyclone V (5CGXFC7C7F23C8) FPGA.

### 5.1 Emergency diesel generator

A classic example that illustrates model-based control functional logic for the EDG, published in Electric Power Research Institute (EPRI) technical report, is illustrated in Fig. 4. The EDG receives a total of fourteen digital input signals and produces two output signals. The output signals are calculated from the input signals using basic combinational logic AND, OR, and NOT operations. The EDG digital control system within a Nuclear Power Plant (NPP) is a safety-critical system required for reactor cooling and other safety functionalities. While the functionality of the EDG is rather simplistic, it is a highly critical system that must be fault-tolerant. To demonstrate the resilience properties of the proposed architecture, the EDG critical application has been implemented on the proposed architecture. This implementation required sixteen functional cells to be connected together in such a way that two critical functions layers are interconnected with each other (see Fig. 4). Each one of the two functional cells (B cells) embedded inside the critical functions layer has to (i) activate a different genetic code (DNA expression) based on the current address of the functional cell and (ii) receive different digital input data through the I/O routing units connected to the input and output ports of the EDG application. When the EDG is subjected to two sequential permeant faults into the functional units of both F0 cell and R0 cell in the critical function layer, the EDG application is healed against the first fault by time 345 ns and the second fault by time 455 ns. This about an 82% increase in time delay to handle two sequential permanent faults – and this delay remains relatively constant as the number of handled faults increases.

### 5.2 Cruise control system

A classic example that illustrates mode-based control seen in process automation applications is the *automotive cruise control system* (*CCS*) illustrated in Fig. 5. The CCS is a closed-loop control system that keeps the vehicle tracking at a constant speed without depressing the accelerator pedal in spite of the external disturbances. This can be achieved by measuring the vehicle speed, comparing it to the desired speed, and then adjusting the throttle output value based on specific control rules like the Proportional Integral (PI) controller. The CCS receives a total of six digital input signals and produces two output signals. The output signals are calculated from the input signals using a combination of some digital control logic and a PI controller.

**Table 1** CCS functionality mapping on 17 functional cells

| Implementation level | Functional cell | Operation |
|---|---|---|
| Level_1 top control logic | FC1, FC2 | NOT, addition |
| | FC3, FC4 | delay, OR |
| | FC5, FC6 | multiplexing, subtraction |
| Level_2 PI controller | FC12, FC13 | multiplication, addition |
| | FC14, FC17 | |
| | FC15, FC16 | comparison, multiplication |
| | FC7, FC8 | multiplexing |
| Level_3 bottom control logic | FC9 | delay |
| | FC10 | addition |
| | FC11 | subtraction |

**Table 2** Operational semantics of CCS application

| Condition | State | Operation |
|---|---|---|
| set | operation is set | target speed = actual speed |
| decrement | speed is decreased | target speed = target speed − 1 |
| increment | speed is increased | target speed = target speed + 1 |
| cancel/brake | speed is cancelled | target speed = 0 |

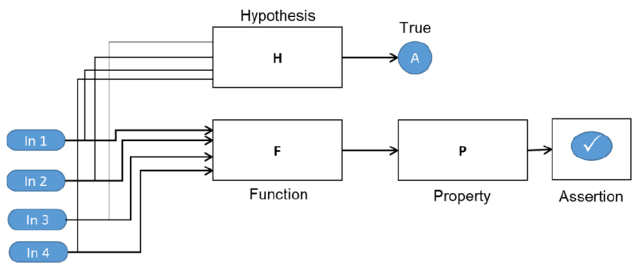

**Fig. 6** *Generic proof structure of Simulink DV*

A block diagram for PI, the controller that is used in many industrial control systems, has been implemented on the proposed hardware architecture with a modest investment in time. To implement the CCS application on the architecture, this application has been partitioned into three levels: level1 (top control logic), level2 (PI controller), and level3 (bottom control logic). Table 1 shows the different operations that are required to perform the mapping process of the CCS application and how they are distributed on 17 functional cells of the proposed architecture, and the four operational semantics of the designed CCS application are shown in Table 2. As a consequence, the functionalities of these three different levels were distributed among three critical functions layers of the architecture illustrated in Fig. 2, and at each layer eight functional cells (designated *F*) are triggered at a specified time.

The experimental simulation results show that the CCS takes at least 35 ns execution time to produce a value '50' for the 'Target' output signal because the clock cycle time is 10 ns. In addition, this signal is generated only by the functional cell FC5, in which the execution of control flow embedded inside the GFB needs only 3.5 clock cycles. Finally, the FC6 output, which represents the error signal based on the difference between the actual speed and target speed, is always connected as an input signal to the PI controller. This controller computes the 'Throttle' output signal value based on the error value to generate the throttle output value at time 430 ns.

The simulation results for the first fault injection case study in which, three sequential transient faults have been injected into the three input registers of FTGFB for the first bio operational-cell (designated *F0* in Fig. 2) in the architecture at times: 180, 240, and 300 ns, hybrid redundancy units are tolerating their impacts quickly, and the output signal is generated without producing any erroneous value at the 'Dat_Out_B' digital output port. This signal represents the result of the GFB executing on the four digital input signals: 'North', 'West', 'East', and 'South'. The second case study shows that two sequential permanent faults have been injected into the GFB units of two cells: B cell (designated *F0* in Fig. 2) and T cell (designated *R0* in Fig. 2). In all cases, faults are identified by embedded self-checking units, and the system is repaired successfully. This type of multiple fault scenario typically occurs in industrial automation systems when there is a cascading disturbance effect due to power fluctuation, electromagnetic interference (EMI), and a latch-up.

## 6 Design practices and verification methods

From the outset, we adopted a model-based design perspective for this work. Model-based design is a design method that establishes a useful framework for the development and integration of a formal executable model system and its environment early in the design cycle. We chose the MathWorks Simulink Toolchain for the presented project. The main tools we used for the design were:

- Simulink and Simulink Stateflow.
- Simulink Design Verifier.
- Simulink HDL coder (automatic code generation).
- Altera Quartus for the analysis, simulation and synthesis of HDL designs.

A critical tool in our verification scheme is the usage of the Simulink Design Verifier (DV) toolbox. DV is a formal verification tool combining both model checking and limited automatic theorem proving [26]. Model-checking, as the name implies, given a model of a system checks to whether this model meets a given property specification. Usually this consists of exploring all states and transitions in the state model. While model checking is finite in nature, the number of states that can be efficiently searched is enormous – making it practical and applicable to real systems. If properties hold, the model checker outputs a confirmation. If a property fails to hold for some possible event sequences, the tool produces counterexamples, i.e., traces of event sequences that lead to failure of the property in question [27].

In Simulink DV, a proof objective is generally specified, as illustrated in Fig. 6. We have a function F for which we would like to prove a certain property *P*. As shown in Fig. 6, the output of function *F* is specified as input to block *P*. Property *P* is a predicate, which should always return true when hypotheses *H* set on the input data flows of the model are satisfied. *P* is therefore connected to an *Assertion* block, while *H* is connected to a *Proof Assumption block*. Whenever an *Assertion* block is used, DV attempts to verify whether its specified input data flow is always true. *Proof Assumption* blocks have the purpose of constraining the input data flows of the model during proof construction. *Proof Assumptions* blocks are not always required, especially if input space does not need constrained [28].

Simulink DV has been used to check a number of properties for the architecture and its two applications. We show one example of functional property proving and the verification model for the FTGFB for one specific functional property proof (see Fig. 7). Specifically, a functional specification in English for the proper sequencing of FTGFB is given as:

> 'If four digital input data lines of the FTGFB are read at the same time in parallel while state machine-based control flow is triggered, the Data-out-B signal will always produce a correct value with the rising edge of the done signal'.

This functional specification is transformed into DV property model and the formal temporal logic expression of this requirement is

$$G(P1 \wedge P2 \wedge P3 \wedge P4 ==> F(Q))$$

where $P1$, $P2$, $P3$, $P4$ = four digital input lines, $G$ is universal quantification of the expression.

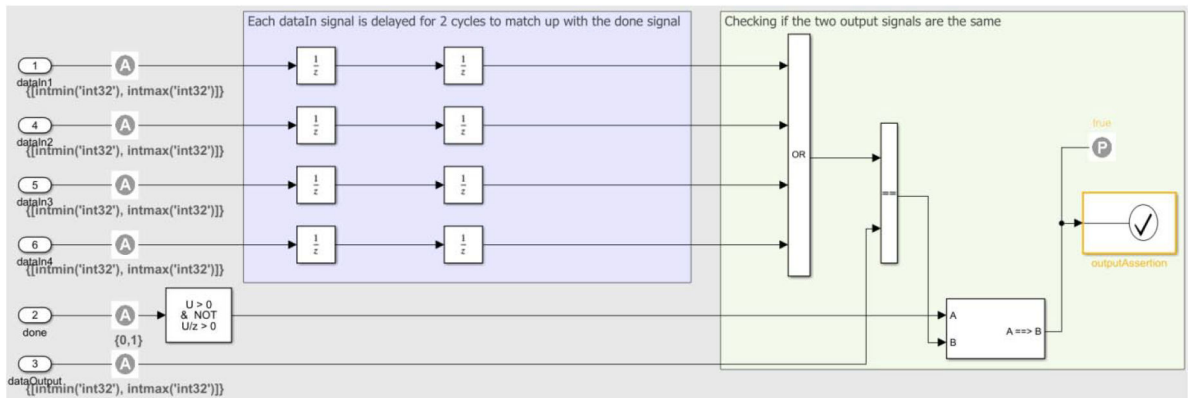

**Fig. 7** *Functional property proving*

**Table 3** Self-healing capacity coverage and area overhead comparison in the EDG implementation with $N*N$ array of cells

| Architecture | Biological concept | Functionality | Advantages and disadvantages | No. of F cells | No. of spare cells | No. of re-routing cells | Self-healing capacity coverage | Area over-head |
|---|---|---|---|---|---|---|---|---|
| proposed self-healing architecture | embryonic cells + immune system + DNA expression | 61131-based unit capable of implementing one function of up to N variables | (1) tolerating transient faults in the registers and the defected cell can be used for unlimited number of times (2) healing against permeant faults in 1131-based unit (3) recovery time is demonstrated | $N*N/2$ | $(N*N)/2 + (N*N)/4$ | — | 1 | 150% |
| re-routing self-healing (Lala) elimination strategy | immune system | LUT-based can implement any function of up to three variables | (1) healing against transient faults in the contents of RAMs (2) cells can tolerate only one fault (3) system cannot use the defected cell for the second time (4) (tolerates only one fault for each f cell) 100% if no. of faults = no. of spares = 8 (5) 66% if no. of faults = 12 > no. of spares (6) recovery time is not verified | $N*N/2$ | $(N*N)/4$ | $(N*N)/2$ | 0.333 | 150% |
| gene control (Yang) elimination strategy | endocrine cellular communication | LUT-based | (1) monitoring and detecting the soft errors only inside the gene memory (2) re-routing with, but not after cell replacement (3) proposed to be used in outer space or deep sea (4) four sequential permanent faults for one working cell and two simultaneous faults in two cells | $N*N/2$ | $(N*N)/2$ | — | 0.666 | 100% |
| voting-by-majority TMR elimination strategy | paralogous gene regulatory circuits | LUT-based | (1) five permanent faults and unlimited number of transient faults in a single working cell with time delay reconfiguring four spare cells and one redundant cell | $N*N/2$ | $(N*N)/2 + (N*N)/8$ | — | 0.833 | 125% |

## 7 Comparative self-healing capacity and area overhead assessment

In this section, we perform a qualitative comparison of the proposed architecture and comparable systems found in the literature. These reference cases include a voting-by-majority Triple Modular Redundancy, self-healing architecture, the re-routing self-healing architecture by Lala *et al.*, and the self-healing architecture inspired by the endocrine cellular communication by Yang *et al.* [7, 8, 29, 30]. Table 3 summarises the comparison results when implementing the EDG application with an array of $N*N$ functional cells. In the comparison table (see Table 3), all the functional cells that are used for either the rerouting purposes or the recovery processes were considered as an additional hardware

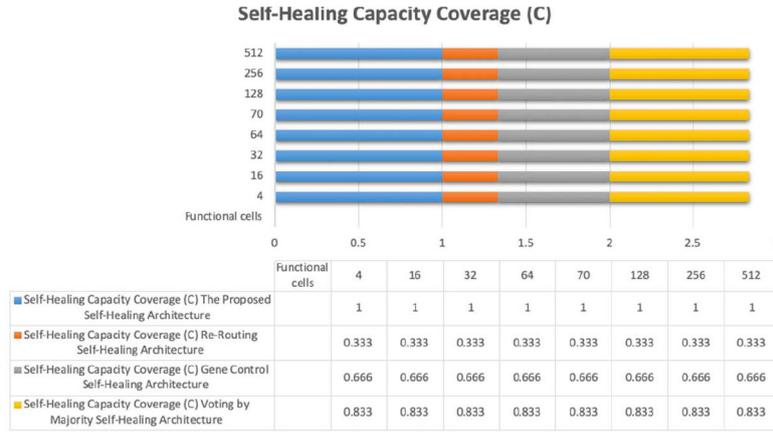

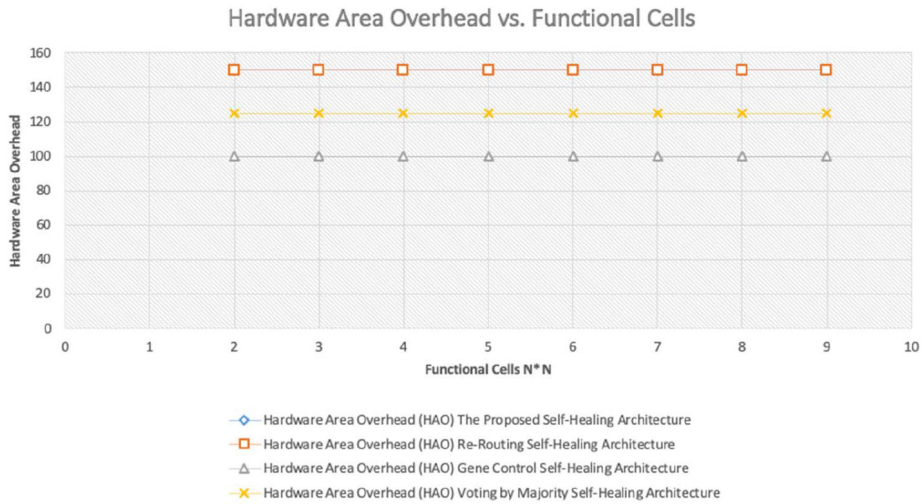

**Fig. 8** *Comparative assessment results of the proposed architecture and comparable systems*
*(a)* Self-healing capacity coverage between different architectures as system size increases, *(b)* Hardware area overhead between different architectures as system size increases overhead. In addition, the maximum number of defective functional cells that can be healed against a number of sequential or concurrent permanent faults has been used in the calculation of the self-healing capacity coverage (*C*). Each of the four self-repairing strategies has a different cell replacement and rerouting process in such a way that each one has different advantages and disadvantages which are presented in Table 3. The self-healing capacity coverage (*C*), is calculated based on the following formula:

$$\text{Self healing capacity coverage}(C) = \frac{\text{SCs}}{\text{SPF}} \quad (1)$$

where SCS represents the total or available number of spare functional cells that can be used for self-healing at an instant in a given time. SPF represents the maximum number of sequential or concurrent permanent faults occurrences that may defect the self-healing architecture at a given time. In relation to the self-healing capacity coverage, we assume up to a maximum of 12 fault occurrences of fault type permanent (SPF = 12) that can impact the self-healing architecture sequentially at different times or concurrently at one instant of time. The self-healing capacity coverage (*C*) has been calculated for the proposed architecture and compared to the other three architectures, as it is shown in Fig. 8*a*.

These results show that the proposed architecture has the potential to achieve high self-healing capacity coverage (*C*) (approaching 1), and as much as the coverage of both self-repairing architectures, the voting-by-majority, and the gene control architecture by Yang as the system size increases. In addition, our proposed architecture requires 150% hardware overhead, and this overhead is approximately equal to the overhead was consumed by the voting-by-majority elimination architecture. The gene control self-repairing architecture requires 100% hardware area overhead and the re-routing self-healing architecture consumes 150%, as shown in Fig. 8*b*. The hardware area overhead that is shown in Table 3 was calculated based on (4)–(6) is considered efficient for our architecture when compared to the re-routing self-healing architecture approach, and the self-healing capacity coverage is considered high

$$\text{Area overhead} = \frac{(\text{No . of spare cells} + \text{No . of routing cells})}{\text{No . of functional cells}} * 100 \quad (2)$$

The reason behind that pre-generated T cells are distributed throughout the system structure in such a way that each functional B cell has its own T cell. As a result, no row or column elimination strategy is needed to recover the system against the failure, which is considered an inefficient method in terms of hardware area resources in other systems. When a cell goes faulty, only one of its surrounding redundant T cells can replace it in a self-healing system. However, as a drawback, the unutilised hardware resources embedded in the same cell when the faulty cell is removed from the system due to only a single fault occurs in one component is considered inefficient.

Regarding the re-routing architecture, self-healing properties can be achieved for the same number of faulty cells defected by permanent faults as a comparison with the three presented architectures. However, the self-healing capacity coverage is less and equal to 0.333 (see Table 3) due to the availability of only four spare cells in a network comprised of 16 functional cells. The results presented in Table 3 are dependent on fault classes that they were assumed, and the proposed architecture was designed for-some other self- healing hardware architectures work only for tolerating transient faults, soft errors, intermittent faults, etc. Consequently, a comprehensive comparison table between the proposed self-healing architectures found in the literature is challenging. However, we can that the proposed architecture is superior to other research works in terms of predicting multiple fault classes, self-healing coverage, and area overhead.

## 8 Conclusion and future work

This paper has presented the architecture, design, and application of a new biological inspired self-healing hardware machine. This machine is intended for the design of CPSs where; (i) resilience to failures and fault tolerance are important, (ii) efficiency in the hardware overhead is needed, and (iii) flexibility to reconfigure and accessible to automation I&C engineers. The adoption of Function Block Programming allows existing PLC applications to be translated into the proposed system more easily and promotes understanding by the automation community. The proposed architecture is a unique approach in that resilience principles are derived from a heterogeneous perspective-combining concepts from both biologically inspired self-healing attributes and system organisation properties to achieve efficient and effective fault tolerance to multiple classes of faults. This architecture has been demonstrated on several systems to date with confirmatory evidence that the approach is feasible and is practical.

Immediate future work will include several important research tasks necessary for translating the research into the pre-commercial evaluation. The first step is to select and develop a means to network the architecture modules to facilitate distributed autonomous control. Second, conduct a large-scale fault injection campaign to collect fault handling data. Third, critical components of the architecture will be formally verified to verify both the data flow and the control flow at runtime for the FTGFB. As a final observation we note the experience of designing and using a self-healing hardware architecture has yielded more information than just quantifying the self-healing and resiliency aspects of the system. The process itself was an iterative learning experience, allowing circumspection into how Bio-inspired systems can be pragmatic solutions to achieving more dependable systems. Therefore, with this project we have attempted to bring new insights into the design of bio-inspired autonomic systems for a range of stakeholders from automation to the Internet of Things.